\documentclass[3p,sort&compress]{elsarticle}

\usepackage{hyperref}
\usepackage{amsmath}
\usepackage{comment}
\usepackage{natbib}
\usepackage{subscript}
\usepackage{xcolor}

\newcommand{\mtr}[1]{\mathbf{#1}}

\journal{Journal of Computational Physics}

\begin{document}

\begin{frontmatter}

\title{A fast moving least squares approximation with adaptive Lagrangian mesh refinement for large scale immersed boundary simulations}

\author[add1]{Vamsi Spandan}
\address[add1]{Physics of Fluids and Max Planck-University of Twente Center for Complex Fluid Dynamics, University of Twente, Enschede, PO Box 217, 7500 AE, Netherlands}
\author[add1,add2]{Detlef Lohse}
\address[add2]{Max Planck Institute for Dynamics and Self-Organization, 37077, G\"ottingen, Germany}
\author[add3]{Marco D de Tullio} 
\address[add3]{Politecnico di Bari, Via Re David 200, 70125, Bari, Italy}
\author[add1,add4]{Roberto Verzicco}
\address[add4]{University of Rome 'Tor Vergata', Via del Politecnico, Rome 00133, Italy}


\begin{abstract}
In this paper we propose and test the validity of simple and easy-to-implement algorithms within the immersed boundary framework geared towards large scale simulations involving thousands of deformable bodies in highly turbulent flows. 
First, we introduce a fast moving least squares (fast-MLS) approximation technique with which we speed up the process of building transfer functions during the simulations which leads to considerable reductions in computational time. We compare the accuracy of the fast-MLS against the exact moving least squares (MLS) for the standard problem of uniform flow over a sphere.  
In order to overcome the restrictions set by the resolution coupling of the Lagrangian and Eulerian meshes in this particular immersed boundary method, we present an adaptive Lagrangian mesh refinement procedure that is capable of drastically reducing the number of required nodes of the basic Lagrangian mesh when the immersed boundaries can move and deform. Finally, a coarse-grained collision detection algorithm is presented which can detect collision events between several Lagrangian markers residing on separate complex geometries with minimal computational overhead. 
\end{abstract}


\end{frontmatter}

\section{Introduction}

Numerical simulations of flows around moving and deformable bodies with complex geometries have been an immense challenge to the fluid dynamics community for a long time. Computational studies using body-fitted methods where the mesh needs to be adjusted according to the instantaneous shape and position of the immersed body become not only mathematically complex but also terminally expensive \cite{dowell2001modeling}. This makes non body-fitted methods such as the immersed boundary method (IBM) very attractive to deal with complex flow systems especially when there is coupling between a carrier fluid and a deformable interface, membrane or a structure \cite{mittal2005immersed}. IBM uses a single time-invariant mesh for the solution of the Navier-Stokes (NS) equations and a separate set of nodes distributed on the surface of an immersed body. The influence of an immersed rigid or deformable body on the surrounding fluid is achieved through a spatially and time varying forcing function in the NS equations. This method was first used by Peskin \cite{peskin1972flow} to study flow around heart valves and has gained immense popularity over the last two decades due to its ability to handle a large class of problems; for example suspension of rigid spheres in turbulent flows \cite{uhlmann2014sedimentation,picano2015turbulent,prosperetti2015life,
fornari2016sedimentation}, deformable liquid-liquid interfaces and elastic membranes \cite{detullio2016moving,spandan2017parallel}, flow around bluff bodies \cite{iaccarino2003immersed} etc. One particular field where IBM has excelled is in the simulation of dispersed rigid particles in highly turbulent flows \cite{uhlmann2014sedimentation,picano2015turbulent,prosperetti2015life,
fornari2016sedimentation}. Recent works have shown that while the extension from rigid particles to deformable particles is highly non-trivial it can be achieved with the help of optimised deformation and parallelisation algorithms \cite{schwarz2016immersed,detullio2016moving,spandan2017parallel}. 

Fully resolved simulations of large scale multi-phase flows are useful not only to understand the fundamental features of the flow but also to benchmark and improve lower-order models which are used extensively in industry. Simulations of such systems which were not possible a decade ago are now being made possible with the combination of a variety of ingredients such as efficient NS solvers for the fluid-phase, IBM procedures with the ability to handle deformable bodies and large-scale parallelisation of the complete solver. However, there is still a need to optimise several sub-components of the resulting software such as interpolation techniques, collision detection algorithms, mesh refinement, load balancing for parallel simulations etc. which are some of the major bottlenecks in computing long-time statistics for {\it large scale} simulations involving IBM (the term {\it large scale} here arbitrarily refers to numerical simulations involving $O(10^4)$ fully resolved deformable bodies with $O(10^3)$ Lagrangian markers each in a turbulent flow on a Eulerian mesh of O($10^9$) or more nodes).  
In this paper, we tackle some of these bottlenecks and introduce simple and effective optimisation algorithms within the immersed boundary framework which are geared towards scaling up simulations of highly turbulent flows with thousands of fully resolved deformable particles. 

As mentioned before, IBM requires two different meshes; an Eulerian one for the solution of NS equations and a Lagrangian mesh on the surface of the immersed body which is primarily used to enforce the interfacial boundary condition. When the immersed bodies can deform, the Lagrangian mesh also serves as the computational grid for the discretisation of the structure equations; such as the Cauchy-Navier equations in elastic deformation problems. While there are multiple variants of IBM which vary in how the influence of the immersed body is imposed on to the carrier fluid, the most popular approach in the case of moving bodies is the one introduced by \cite{uhlmann2005immersed}. In this approach, the forcing function on each Lagrangian node is computed from a {\it cloud} of surrounding nodes of the Eulerian mesh. Since the Lagrangian nodes do not coincide with the Eulerian mesh at every time instant, interpolation and extrapolation algorithms are implemented into the flow solver; this brings into play a couple of issues. Firstly, it is important to ensure that the transfer functions used to exchange information between the Eulerian and Lagrangian meshes conserve momentum and energy between the immersed body and the surrounding fluid in addition to maintaining a specific order of accuracy compatible with the flow solver. Secondly, the calculation of the transfer function should not become computationally intensive as this hampers the ability of the flow solver to handle {\it large scale} systems. Another major issue in numerical simulations involving IBM is the coupling or  dependence of the resolution of the Lagrangian mesh with the Eulerian mesh. According to the standard implementation there must be, at least, one-to-one correspondence of Eulerian and Lagrangian nodes if we have to prevent {\it holes} or {\it flux leakage} at the immersed surface. In other words, having a Lagrangian resolution coarser than the Eulerian counterpart results in a discontinuous representation of the immersed boundary which leads to artificial effects in the flow. Typically, the acceptable choice is that the length scale of an individual Lagrangian element is set to 0.7-0.9 times that of the local Eulerian grid size. However, it is important to note that one-to-one correspondence of Eulerian and Lagrangian nodes becomes terminally expensive in wall-bounded flows where the mesh is refined in the near wall region and this forces the Lagrangian mesh to be unnecessarily refined when the moving immersed body evolves in the bulk region. 

In recent works, the moving least squares (MLS) approximation has been used to build the transfer functions between the Eulerian and Lagrangian meshes and in particular it has been favoured for the simulation of deformable interfaces and membranes \cite{vanella2009moving,detullio2016moving,spandan2017parallel}. MLS allows for interpolating both the value and the spatial derivatives of a flow variable at any given point based on a set of field nodes defined within a local support domain in addition to ensuring the transfer function is smooth in presence of moving and deforming boundaries. MLS was first introduced by mathematicians to solve the problem of data fitting and surface reconstruction and given its highly generic nature it has been used in several fields thereafter \cite{belytschko1994fracture,belytschko1996meshless,krongauz1996enforcement,hegen1996element,atluri1999analysis,schaefer2006image,fleishman2005robust,kolluri2008provably,zeng2004curve,kobbelt2004survey,vanella2009moving,detullio2016moving,spandan2017parallel}. While MLS provides several advantages in handling moving and deformable bodies in inhomogeneous unsteady flows, there are certain inherent issues in the algorithm which make it difficult to use. In comparison to conventional interpolation techniques, MLS is a computationally intensive algorithm as it requires the construction of multiple weight matrices and a matrix inversion 
for every Lagrangian marker where the approximation is required. This severely hampers the wall-clock time of the simulation when there are $O(10^6-10^7)$ Lagrangian elements immersed in the flow. In the first part of this paper, we introduce a simple algorithm (referred to as fast-MLS) where we compute the MLS shape functions only once at the beginning of the simulation at pre-defined nodes and then use a relatively inexpensive interpolation technique (Shepard's interpolation here) to estimate the shape function at any position. With this we aim to achieve the smoothness and accuracy of MLS shape functions while eliminating the operation costs of constructing and inverting weight matrices as in the exact-MLS approach. We compare the results and the total computational time from both approaches for the well-studied problem of uniform flow over a stationary sphere.  

As mentioned earlier, due to the coupled nature of the Eulerian and Lagrangian meshes in IBM simulations, any increase in the Eulerian mesh forces an increase in the Lagrangian mesh. This can severely hamper the performance of the code when the immersed bodies are moving and the Lagrangian mesh has to be evolved in the coarse as well as in the refined Eulerian mesh regions. In order to tackle this problem, we propose segregating the Lagrangian mesh into two parts (i) base mesh which can be used to deform the immersed body (ii) refined mesh for computing the forcing function which is required to satisfy the interfacial boundary condition. The base mesh needs to be initialised only once during the start of the simulation while the refined mesh is defined dynamically according to the local Eulerian resolution. This approach not only makes the flow solver more flexible since the deformation governing equations can now be discretised independently of the Eulerian mesh, but it also reduces the memory load of the code as well as the computational costs. To test the validity of the proposed approach, we compare results of a deforming drop in a cross-flow by both fully resolving the Lagrangian mesh according to the Eulerian resolution and also using the refinement technique which requires relatively less operations and memory. 

Finally, when simulating thousands of deformable bodies dispersed in a turbulent flow, detecting and modelling collision between them becomes a challenging computational task. To this extent, in the last part of the paper, we discuss a simple and easy-to-implement coarse grained collision detection algorithm which can be used to detect collision events between markers residing on different deformable immersed bodies with minimal computational overhead. It is worth to stress that, our primary focus is on collision detection between deformable bodies within the immersed boundary framework and not on the specific physical collision modelling that can be parametrised by separate models. 
 
The paper is organised as follows. In the next section we describe the fast-MLS approach and compare the results from this approach with the conventional MLS-IBM simulations. In section 3, we describe the adaptive Lagrangian mesh refinement technique and compare the results from this technique with previous works. In section 4, we discuss the details and results from the coarse-grained collision detection algorithm and then provide a summary and outlook in section 5.  

\section{Fast moving least squares for the immersed boundary method}

First, we describe briefly the MLS approximation used in the context of IBM \cite{vanella2009moving,detullio2016moving,spandan2017parallel}. In MLS, a field variable $q$ given on a pre-defined set of nodes is approximated at any desired point lying within these nodes and positioned at $\mtr x_l$ as follows:

\begin{equation}
Q(\mtr x_l) = \mtr p^T(\mtr x_l)\mtr a(\mtr x_l)
\label{eqn:mls}
\end{equation}
$Q$ is the approximated quantity while $\mtr p^T(\mtr x_l)$ is a basis function vector with dimension $m$. $\mtr x_l$ is the position vector of the Lagrangian marker $l$. $\mtr p^T(\mtr x_l)=[p_1(\mtr x_l),...,p_m(\mtr x_l)]$ is built using the Pascal's triangle and pyramid in two and three-dimensions, respectively. In this work, we consider a linear basis function with $\mtr p^T(\mtr x)=[1,x,y,z]$, i.e. $m=4$ which has already been shown to be a cost-efficient and reliable choice for flow simulations incorporating the immersed boundary method \cite{vanella2009moving,detullio2016moving,spandan2017parallel}. It is to be remembered that the fast-MLS approach described later is independent of the choice of $m$. 

The vector $\mtr a(\mtr x_l)$ is computed by minimising a weighted L2 norm $J$ which is defined as:
\begin{equation}
J=\sum_{k=1}^{N_e}W(\mtr x_l-\mtr x_k)[\mtr p^T(\mtr x_k)\mtr a(\mtr x_l)-q_k]^2
\label{eqn:mls_j}
\end{equation}
$N_e$ is the number of nodes in the Eulerian grid which are in support of the Lagrangian marker $l$ and the selection of these nodes is described later. $W(\mtr x_l-\mtr x_k)$ is a weight function for point $\mtr x_k$ and can be computed in different ways; for example using exponential functions, cubic splines, quadratic splines etc. \cite{liu2005introduction}. The choice of the weight function is arbitrary and primarily depends on the desired levels of accuracy provided the conditions for continuity and smoothness of the weight function within the support domain are satisfied. A spline weight function with a desired order of continuity can be constructed using the method described in \cite{liu2003smoothed}. Minimising $J$ with respect to $\mtr a(\mtr x_l)$ leads to the relation $\mtr A(\mtr x) \mtr a(\mtr x)=\mtr B(\mtr x)\mtr q$, where $\mtr q$ is a vector containing the values of the field variable $q$ on the pre-defined set of nodes; $\mtr A(\mtr x)$, $\mtr B(\mtr x)$ and $\mtr q$ are defined as:

\begin{equation}
\mtr A(\mtr x_l)=\sum_{k=1}^{N_e}W(\mtr x_l-\mtr x_k)\mtr p(\mtr x_k)\mtr p^T(\mtr x_k)
\label{eqn:mls_A}
\end{equation}
\begin{equation}
\mtr B(\mtr x_l)=[W(\mtr x_l-\mtr x_1)\mtr p^T(\mtr x_1)...W(\mtr x_l-\mtr x_{N_e})\mtr p^T(\mtr x_{N_e})]
\label{eqn:mls_B}
\end{equation}
\begin{equation}
\mtr q=[q_1 ... q_{N_e}]^T
\label{eqn:mls_q}
\end{equation}
Minimising equation (\ref{eqn:mls_j}) with respect to $\mtr a(\mtr x)$ gives the coefficient vector $\mtr a(\mtr x) = \mtr A^{-1}(\mtr x) \mtr B(\mtr x) \mtr q$ and the approximated value of the field variable $q$ at $\mtr x_l$ can be computed as follows:

\begin{equation}
Q=\mtr p^T(\mtr x) \mtr A^{-1}(\mtr x) \mtr B(\mtr x) \mtr q
\label{eqn:mls_ql}
\end{equation}
The approximation in equation (\ref{eqn:mls_ql}) can also be written as $Q=\mtr \Phi^T(\mtr x) \mtr q$, where $\mtr \Phi^T(\mtr x)$ is a vector of shape function coefficients of size $1\times N_e$ and is computed as $\mtr \Phi^T(\mtr x)=\mtr p^T(\mtr x) \mtr A^{-1}(\mtr x)\mtr B(\mtr x)$. This shape function is computed for every Lagrangian marker in the flow and is used to approximate the value of the required Eulerian quantity (typically fluid velocity) at the position of the marker. Approximation of the fluid velocity gives information on the magnitude of the forcing that needs to be transferred to the Eulerian grid which ensures that the interfacial boundary condition is satisfied. If the vector $\mtr q=[u_1 ... u_{N_e}]^T$, $u_k$ being the fluid velocity defined at the Eulerian nodes and its corresponding approximated quantity $Q=U$, the total IBM force corresponding to the Lagrangian marker $l$ is written as $F_l^i=(V_b^i-U^i)/\Delta t$, where $F_l^i$, $V_b^i$ and $U^i$ are the IBM force, desired velocity of the fluid at the Lagrangian marker and the interpolated fluid velocity in the $i^{\text {th}}$ direction. In this way, the actual velocity of the fluid at the marker is forced to be equal to the desired velocity $\mtr V_b$, in order to impose a no-slip condition. Similar strategies can be used to impose a free-slip boundary condition \cite{kempe2015imposing}. Once the forcing has been determined, the next step is to distribute it onto the Eulerian grid in such a way that the total force during the transfer is conserved which leads to the following relation:

\begin{equation}
\sum\limits_{k=1}^{N_e}\mtr f_k\Delta V_k = \sum\limits_{l=1}^{N_l}\mtr F_l\Delta V_l, 
\label{eqn:mls_for}
\end{equation}
$\Delta V_k$ is the volume of the Eulerian cell $k$ and $\Delta V_l=A_lh_l$ is the forcing volume associated with the Lagrangian marker $l$; $A_l$ is the surface area of the Lagrangian marker $l$ and $h_l=\sum\limits_{k=1}^{N_e}\phi_k^l(\Delta x_k+\Delta y_k + \Delta z_k)/3$. Once the shape function is computed, the forcing to be included in the Eulerian cell $k$ is computed as $\mtr f_k=\sum\limits_{l=1}^{N_l}c_l\phi_k^l\mtr F_l$. The scaling factor $c_l$ is calculated such that equation (\ref{eqn:mls_for}) is satisfied and is given as follows:
\begin{equation}
c_l=\frac{\Delta V_l}{\sum\limits_{k=1}^{N_e}\phi_k^l\Delta V_k}     
\label{eqn:mls_cl}
\end{equation}
$\phi_k^l$ are individual elements of the vector $\mtr \Phi^T(\mtr x)$ and represents the shape function value for the Eulerian cell $k$ corresponding to the Lagrangian marker $l$. The construction of the auxiliary matrices $\mtr A(\mtr x)$ and $\mtr B(\mtr x)$ followed by an inversion of $\mtr A(\mtr x)$ are the computationally most intensive parts of the MLS approximation. For stationary bodies, only a single MLS interpolation per Lagrangian marker at the beginning of the simulation is required. However, when the immersed bodies move or deform, in addition to performing the MLS interpolation every time step, we also need to evaluate hydrodynamic forces, interpolating pressure and velocity gradient components at the marker position, which leads to additional MLS operations. Since these steps have to be performed for every individual Lagrangian marker, it results in a steep increase of the computational time with increasing Lagrangian resolution. In the fast-MLS approach we eliminate the process of computing the shape function $\mtr \Phi(x)$ during the simulation and construct it only at the start of every simulation for a pre-defined set of nodes. The shape function for any arbitrary Lagrangian marker is then computed through a relatively inexpensive interpolation procedure such as tri-linear interpolation or the Shepard's interpolation. The procedure for selecting the nodes used for computing the shape functions and interpolation of the shape functions is now detailed. 

\begin{figure}
  \centerline{\includegraphics[scale=1.0]{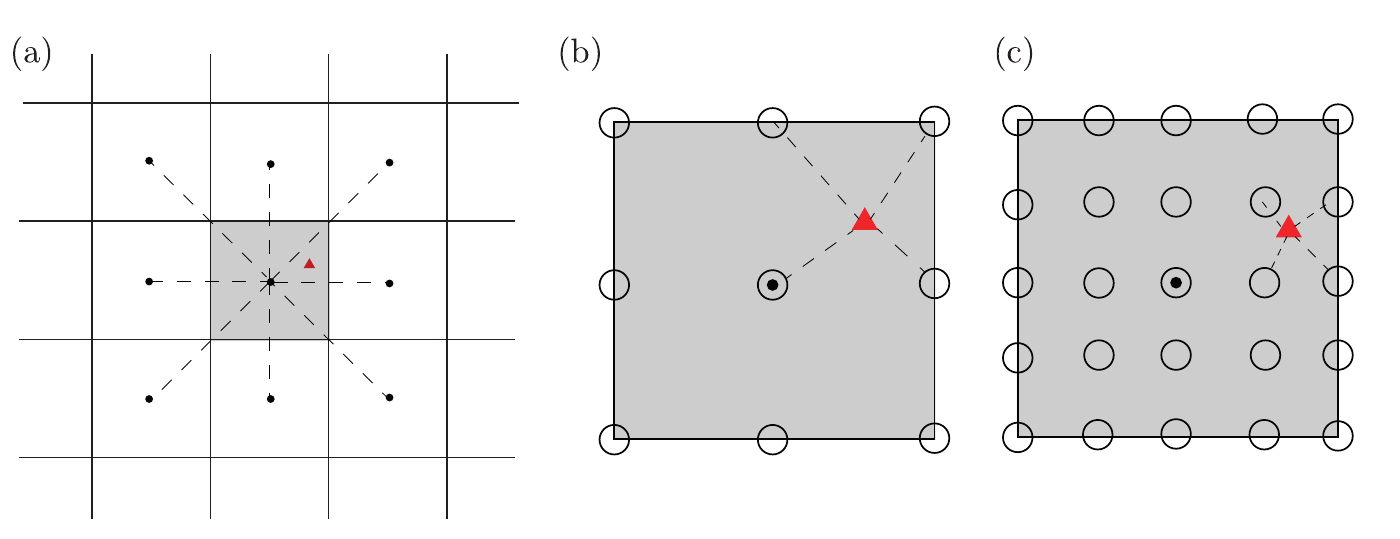}}
  \caption{(a) Schematic of a two-dimensional support domain of the Lagrangian marker (shown by red triangle); dotted lines show all the Eulerian cell centres in support of the Lagrangian marker. (b) Home Eulerian cell of the Lagrangian marker discretised by $N_i=9$ and (c) $N_i=25$ sub-Eulerian nodes, respectively.}
\label{fig:gridsch}
\end{figure}

In figure \ref{fig:gridsch}(a) we show the schematic of a two-dimensional support domain used for an arbitrary Lagrangian marker. For the sake of brevity, in the following, we restrict to a two-dimensional system; although the numerical examples given in the results are all three dimensional flows. The support domain for our exact-MLS approximation in a two-dimensional system consists of $N_e=9$ (3$\times$3) Eulerian cells \cite{detullio2016moving,spandan2017parallel}. The support domain is built by first identifying the home cell of the Lagrangian marker (indicated by the grey region in figure \ref{fig:gridsch}(a)) and then taking its neighbouring cells. In fast-MLS we take advantage of the fact that the shape function of Lagrangian markers at the same relative distances to the Eulerian cell centres of their respective support domains will be exactly the same. While the calculations given below are for a uniform mesh, the same method can also be applied when the mesh is stretched in a single direction, the only difference being a requirement to store additional shape functions at the start of the simulation.  

To incorporate the fast-MLS approach, we first discretise a single Eulerian cell using an imaginary mesh (hereafter called the sub-Eulerian mesh) before the start of the simulation as shown by hollow circles in figure \ref{fig:gridsch}(b); in this case, the resolution of the sub-Eulerian mesh is $N_i=9$ points. Next, we compute the exact shape functions at each of these points using the algorithms described in equations (\ref{eqn:mls_A})-(\ref{eqn:mls_ql}). This leads to nine different exact shape vector functions which contain nine coefficients each. During the simulation, the location of the Lagrangian marker is identified with respect to the sub-Eulerian mesh in its home cell and the shape function is estimated as follows:

\begin{equation}
\mtr \Phi_f(\mtr x_l) = \sum\limits_{i=1}^{4} \alpha_i \mtr\Phi_e^i(\mtr x_i)
\label{eqn:phif}
\end{equation}
\begin{equation}
\alpha_i = \frac{1/d_j^2}{\sum\limits_{j=1}^{4}1/d_j^2}
\label{eqn:alphai}
\end{equation}
$\mtr \Phi_f(\mtr x_l)$ is the shape function estimated from the fast-MLS approach, $\mtr \Phi_e(\mtr x_i)$ is the exact shape function calculated at the nodes of the sub-Eulerian mesh at the start of the simulation, $\mtr x_i$ is the position of the node $i$ on the sub-Eulerian mesh; $\alpha_i$ are the interpolation coefficients computed through Shepherd's procedure which is a relatively inexpensive interpolation technique and depends only on $d_i$ which is the distance between the Lagrangian marker and the nodes of the sub-Eulerian mesh in consideration. The number of Eulerian sub-nodes can be further refined and discretised using 25 points as shown in figure \ref{fig:gridsch}(c) in which case the shape functions on 25 (5$\times$5) different nodes with 9 coefficients each would need to be stored at the start of the simulation. The interpolation of the shape function according to equation \ref{eqn:phif} would then be computed based on a different set of nodes. The next level of refinement of the sub-Eulerian mesh would consist of 81 (9$\times$9) nodes. A comment is necessary regarding the choice of the number of sub-Eulerian nodes used in the calculation of $\alpha_i$ in equations (\ref{eqn:phif}) and (\ref{eqn:alphai}). While estimates from all the sub-Eulerian nodes can be used in the calculation of $\alpha_i$ (for example 9 in figure \ref{fig:gridsch}(b) and 25 in \ref{fig:gridsch}(c)), we find that given the $1/d^2$ dependence of the coefficients, including additional sub-Eulerian nodes in the calculation of $\alpha_i$ does not improve the estimation significantly. 

\begin{figure}
  \centerline{\includegraphics[scale=1.0]{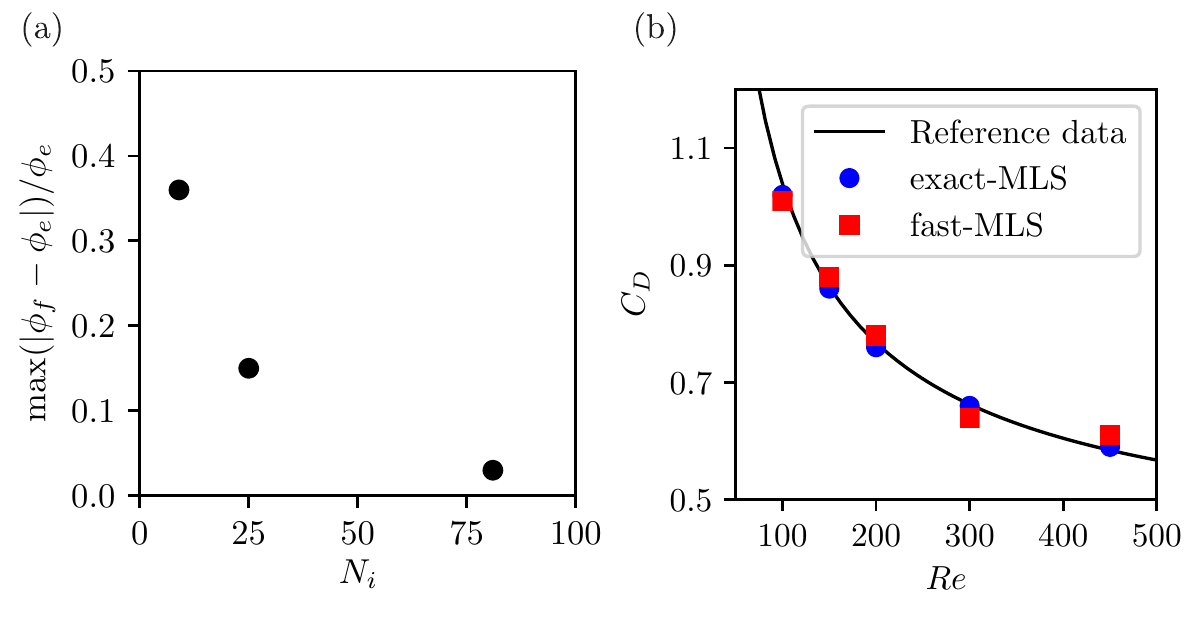}}
  \caption{(a) Maximum relative difference in the shape function coefficients computed from the fast-MLS ($\phi_f$) and the exact-MLS ($\phi_e$) techniques versus the number of grid points used in the sub-Eulerian mesh. (b) Drag coefficient $C_D$ of a stationary rigid sphere facing a uniform flow stream versus the Reynolds number $Re$. In both (a) and (b), the marker size is inclusive of the error-bar. Reference data in (b) is from \cite{abraham1970functional}.}
\label{fig:phicd}
\end{figure}

We now compare the shape function coefficients calculated using both fast-MLS and exact-MLS techniques. In figure \ref{fig:phicd}(a) we show the normalised maximum difference in the shape function coefficients versus the number of grid points used in the sub-Eulerian mesh. In order to ensure, that the maximum relative difference is not affected by position of the Lagrangian marker, we run the tests for 1000 Lagrangian markers randomly positioned in the Eulerian cell. The maximum relative difference is sampled over the 1000 randomly placed Lagrangian markers.
We see that with increasing sub-Eulerian mesh resolution there is a rapid decrease in the difference between the shape functions computed from the exact-MLS and the fast-MLS approaches. While a sub-Eulerian mesh discretised using 9 (3x3) or 25 (5x5) points leads to unacceptable deviations in the shape function coefficients, the error with 81 (9$\times$9) points is already below 3\%. We now use a sub-Eulerian mesh discretised in three-dimensions using 729 (9$\times$ 9$\times$ 9) points in the fast-MLS technique to compute the drag coefficient of a stationary sphere in a uniform flow. The error in the shape function coefficients in a three dimensional system follows the same trend as shown in figure \ref{fig:phicd}(a), while the only difference is the increase in the number of $N_i$, the sub-Eulerian grid nodes. In figure \ref{fig:phicd}(b), we compare the sphere drag coefficient versus the Reynolds number which is defined as $Re=UD/\nu$ ($U$ is the free stream velocity, $D$ is the diameter of the sphere and $\nu$ is the kinematic viscosity of the fluid). A good agreement can be seen between the drag coefficients obtained from the simulations involving the fast-MLS and exact-MLS approach. The scaling factor $c_l$ of equation (\ref{eqn:mls_cl}) which depends on the shape function coefficients estimated from the fast-MLS approach, volume of Lagrangian markers and the corresponding Eulerian cell ensures that there is no loss of momentum while transferring information between the Eulerian and Lagrangian meshes. 

\begin{figure}
  \centerline{\includegraphics[scale=1.0]{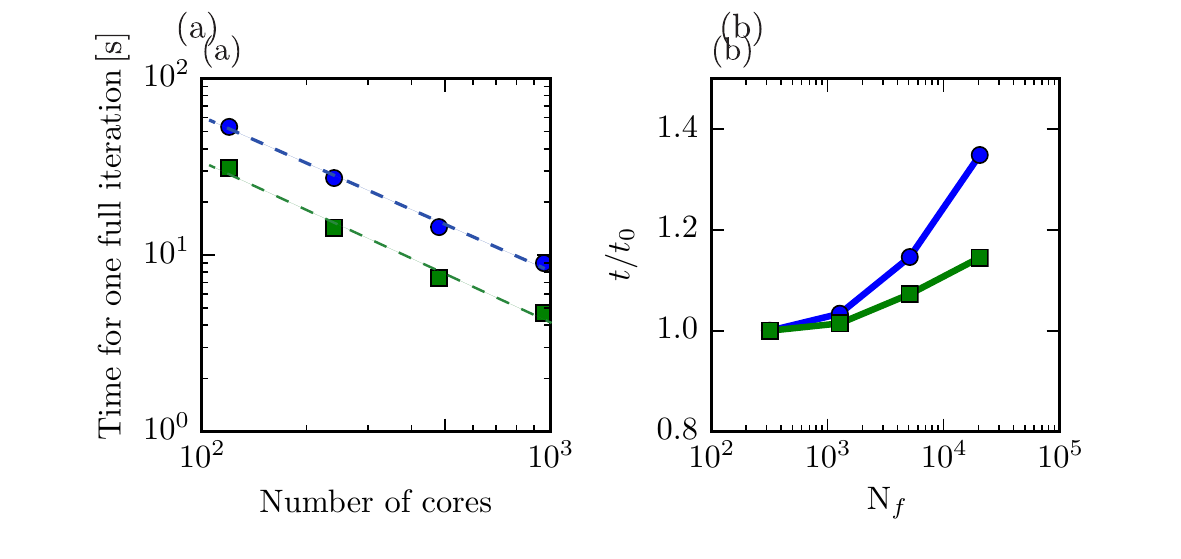}}
  \caption{(a) Scaling plot showing the time step for one full iteration of the solver versus the number of cores used for both the exact-MLS and fast-MLS techniques. (b) Comparison of the total time taken for one full iteration versus the total number of Lagrangian markers immersed in the flow. $t$ is normalised using the time taken for the first data point at $N_f=320$. Circles correspond to exact-MLS while square markers correspond to fast-MLS.}
\label{fig:scale}
\end{figure}

In figure \ref{fig:scale}(a), we show a comparison of the computational performance of the flow solver when using fast-MLS and exact-MLS techniques. The parallelisation algorithms and data structures used in this solver have already been described in \cite{spandan2017parallel}. These simulations were performed on the thin nodes of the Dutch supercomputing facility 'Cartesius' where each node is composed of 2x12 core 2.6 GHz Intel Xeon E5-2690 v3 CPU's. For this test case, we initialise a total of 16000 spherical particles each discretised using 320 faces (i.e. a total of 5 Million Lagrangian markers) in a doubly-periodic Cartesian box (discretised using an Eulerian resolution of 720x720x3840). The flow is driven in one of the periodic directions using a uniform pressure gradient while the immersed particle can move and deform based on the techniques described in \cite{spandan2017parallel}. It can be seen that the fast-MLS approach while ensuring good scalability also reduces the total wall clock time per time step. Here, one full iteration consists of integrating the fluid and immersed body governing equations for one time-step. In order to further analyse the computational boost provided by the fast-MLS approach we now fix the Eulerian resolution and increase the total number of Lagrangian markers ($N_f$) in the flow. For these simulations the Eulerian grid is kept fixed to 120x120x720 and a total of 120 computing processors were used. In figure \ref{fig:scale}(b) we plot the non-dimensional time taken for one full iteration with increasing number of faces where the time is normalised using the time taken for the first data point i.e. $N_\text{f}=320$. Here it is important to point out that for these simulations the immersed bodies can both move and deform according to the deformation algorithm discussed by \citep{spandan2017parallel}. It implies that without the fast-MLS approach  a single iteration would consist of multiple exact-MLS operations. The gain in computational time achieved with the fast-MLS approach can be seen clearly from figure \ref{fig:scale}(b). 

A comment is necessary on the memory requirements of the sub-Eulerian cell approach within the fast-MLS formulation. The memory load is manageable for two reasons. First, in the case of a uniform mesh, the sub-Eulerian cell approach takes advantage of the fact that the shape functions depend entirely on the relative position of the Lagrangian marker with the neighbouring Eulerian cells. Thus, only the shape coefficients of the sub-Eulerian cells of a single Eulerian cell will need to be stored by all processors. Second, in the case of non-uniform grid stretching in one direction, the coefficients of sub-Eulerian cells in the direction of the stretched grid is stored. Although this increases the memory requirement in comparison to a case with uniform grid spacing, for a simulation with multiple processors, each processor will only need to store the information of the Eulerian cells allocated to its memory and not of all the Eulerian cells in the stretched direction. The memory requirement for the largest cases varies from O(1)MB for uniform grid spacing to O(10)MB for non-uniform grid spacing, which is easily manageable on modern computing architectures.

\section{Decoupling the Lagrangian meshes}

\begin{figure}
  \centerline{\includegraphics[scale=1.0]{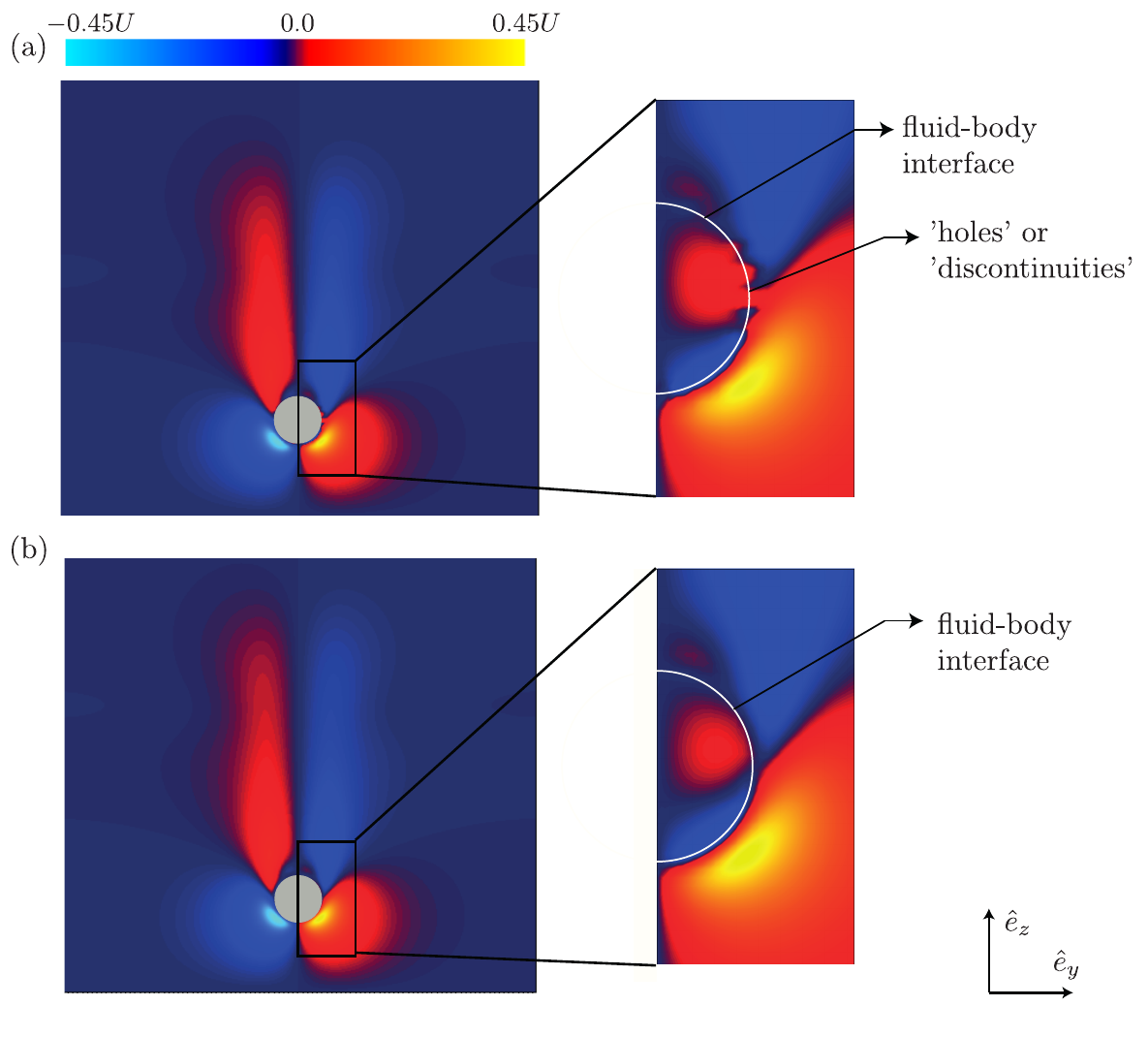}}
  \caption{Flow over a stationary rigid sphere at $Re=100$ for (a) Coarse Lagrangian mesh (b) Refined Lagrangian mesh. The flow is in $\hat e_z$ direction. The right panels show a zoom of the flow around the sphere where 'holes' or 'discontinuities' in the flow can be easily detected in the case of a coarse Lagrangian mesh. The colormap represents the velocity in the $\hat e_y$ direction while $U$ is the ambient flow stream in $\hat e_z$ direction}
\label{fig:fholes}
\end{figure}

We now move on to describing the technique of adaptive Lagrangian mesh refinement and how it can be used in simulations involving IBM. The purpose of the refinement is to tackle two challenges in IBM simulations. Firstly, in MLS-IBM simulations, the resolution of the Lagrangian mesh is inherently coupled to the Eulerian mesh due to the continuous transfer of information between both meshes. If the Lagrangian mesh is too coarse, the Eulerian flow field is not forced uniformly which leads to discontinuities (often called 'holes') in the interfacial boundary condition. This is illustrated in figure \ref{fig:fholes}, where we simulate flow over a stationary rigid sphere in two different cases (i) coarse Lagrangian mesh i.e. $l/h\sim 2.0$, and (ii) refined Lagrangian mesh, $l/h\sim 0.8$, where $l$ is the mean edge length of the triangular elements and $h$ is the mean grid spacing in the Eulerian mesh. It can be clearly seen from the right panel of figure \ref{fig:fholes}(a) that when the Lagrangian discretisation is coarser than the local Eulerian one, the flow develops discontinuities at the interface between the immersed body and the fluid. In contrast, when the Lagrangian mesh is sufficiently fine (as seen in figure \ref{fig:fholes}(b)), i.e. the mean edge length of the triangular markers is smaller than that of the local Eulerian grid spacing, the flow is smooth and the integrity of the interface between the immersed body and the fluid is maintained  

Tackling the problem of discontinuity in the flow with moving immersed bodies using a brute-force approach would mean resolving the Lagrangian mesh according to the smallest Eulerian grid cell, since the position of these bodies is not known in advance and they could end up in the finest mesh regions. A Lagrangian resolution as fine as the smallest local Eulerian mesh solves this problem but this leads to inclusion of additional Lagrangian markers which become redundant when the immersed body evolves in regions where the Eulerian mesh is coarse. An immediate consequence is a massive increase of the total computational time. Secondly, if the immersed body can also deform, the same Lagrangian mesh that is used for enforcing the interfacial boundary condition (IBM mesh) is used for the structural solver. This implies that a refinement in the IBM mesh automatically implies a refinement in the structural solver mesh which may lead to unnecessary computational nodes for discretising the deformation governing equations and small time steps to track the dynamics of small structural elements. Here, we focus on describing the refinement procedure and the results from the simulations of a deforming drop with and without such refinement.  

\begin{figure}
  \centerline{\includegraphics[scale=1.0]{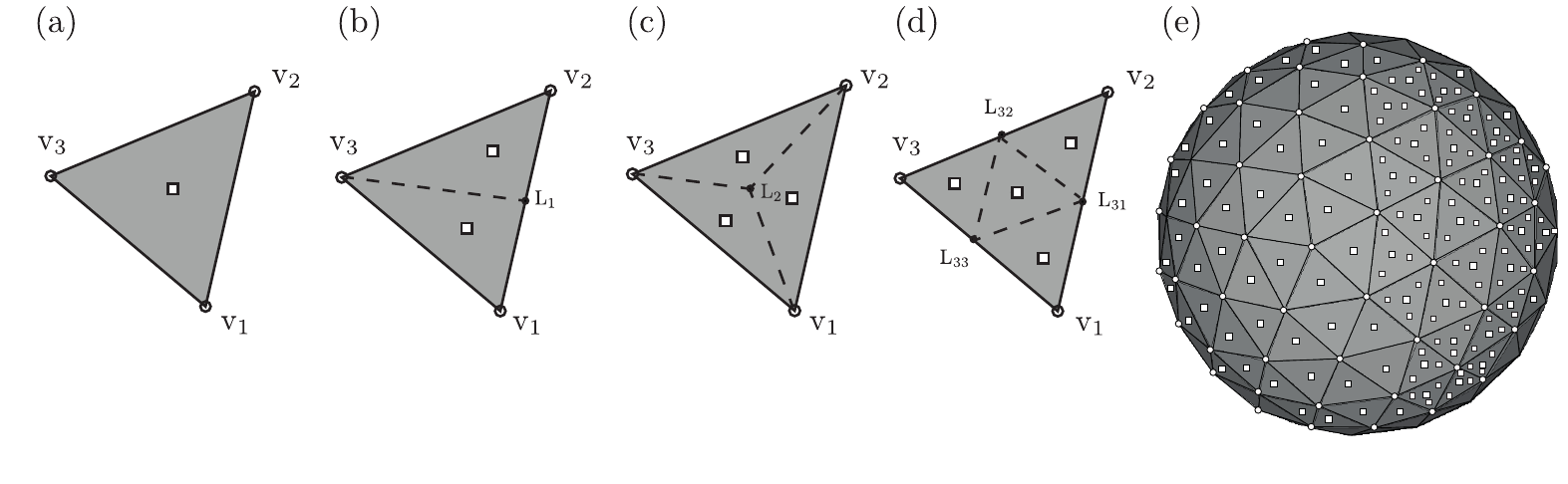}}
  \caption{Triangulations for the adaptive Lagrangian mesh refinement. (a) initial base triangular element. (b), (c), (d) show the first, second and third level of refinements, respectively. (e) A sphere where the right half of the sphere is refined using the third level while the left half still has its initial base mesh.} 
\label{fig:triangs}
\end{figure}

The basic idea of the Lagrangian mesh refinement is shown in figures \ref{fig:triangs}(a)-(d). Consider a triangular element as shown in figure \ref{fig:triangs}(a) which is composed of the three vertices ($\mtr V_1, \mtr V_2, \mtr V_3$) and a centroid (represented by a square symbol). Given the vertices of any triangular element, the position of the centroid can be directly computed as $\mtr C=(\mtr V_1+\mtr V_2+\mtr V_3)/3$ We can further refine the triangular element in multiple levels as shown in figures \ref{fig:triangs}(b)-(d). In the first level of refinement (figure \ref{fig:triangs}(b)), the triangle is split into two such that $\mtr L_1=0.5(\mtr V_1+\mtr V_2)$. In the next level of refinement (figure \ref{fig:triangs}(c)) the triangular element is split in such a way that $\mtr L_2$ is at the centroid of the triangle element i.e. $\mtr L_2=(\mtr V_1+\mtr V_2+\mtr V_3)/3$. Similarly the triangle can be further refined as shown in the figure \ref{fig:triangs}(d); $\mtr L_{31}=0.5(\mtr V_1+\mtr V_2$), $\mtr L_{32}=0.5(\mtr V_2+\mtr V_3)$, $\mtr L_{33}=0.5(\mtr V_3+\mtr V_1)$. Here it is important to note that such a refinement procedure is computationally inexpensive and for a given base mesh, the positions of the vertices and centroids of the refined triangles can be easily computed dynamically during the simulation. The refinement procedure described above holds when the \lq mass\rq is distributed uniformly over the triangular network, while in case on a non-uniformly distributed mass, appropriate weights have to be given to the individual vertices. The locally refined Lagrangian markers take part only in enforcing the interfacial boundary condition but not in the deformation of the immersed body, i.e. no additional curvature information is added to the body. Thus, if the initial Lagrangian resolution is already sufficient to accurately represent the surface during deformation, the decoupling ensures no redundant calculations are performed for the structural elements.  

\begin{figure}
  \centerline{\includegraphics[scale=1.0]{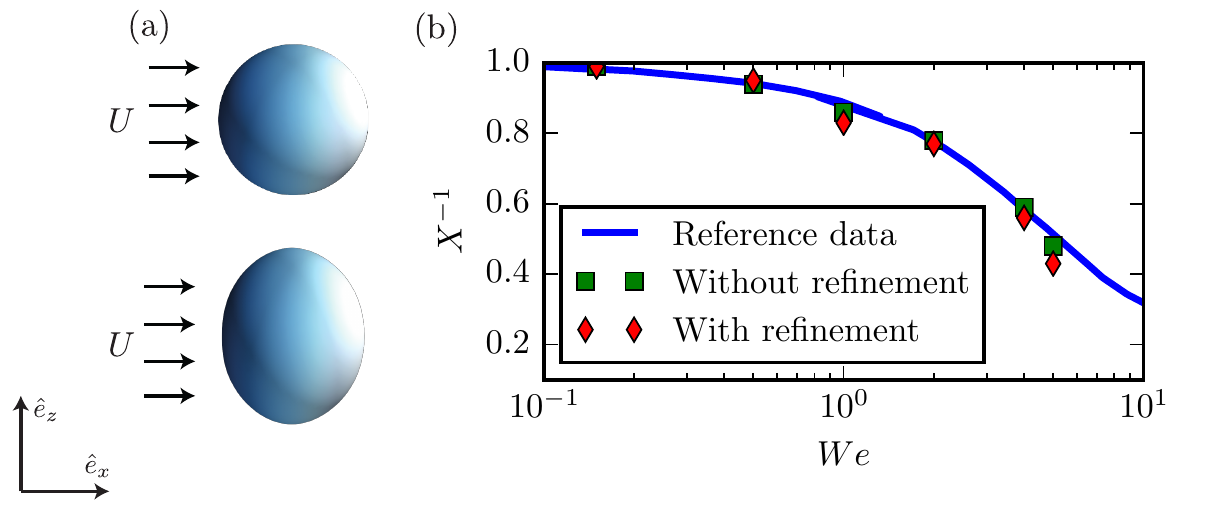}}
  \caption{(a) Snapshots of a drop immersed in a cross flow at two time instants; top and bottom panel corresponds to its initial and deformed state, respectively. (b) Comparison of the inverse of the mean aspect ratio of the drop versus the Weber number with and without the refinement technique. The reference data is given by a fit from several experimental measurements given in \cite{loth2008quasi}.}
\label{fig:lcomp}
\end{figure}

\begin{table}
  \begin{center}
\def~{\hphantom{0}}
	\begin{tabular}{c|c|c}
	$We$ & Base $N_f$ & Refined $N_f$ \\
	\hline
	\hline
	0.15 & 162  & 324    \\
	0.50 & 162  & 628    \\  
    1.00 & 642  & 1926   \\
    2.00 & 642  & 2568   \\
    4.00 & 2562 & 7686   \\
    5,00 & 2562 & 10248  \\
    \hline
    \end{tabular}
    \caption{Details on the base mesh and refined mesh of the Lagrangian surface for the simulations shown in figure \ref{fig:lcomp}}
    \label{tab:wemesh}
  \end{center}
\end{table} 

\begin{figure}
  \centerline{\includegraphics[scale=1.0]{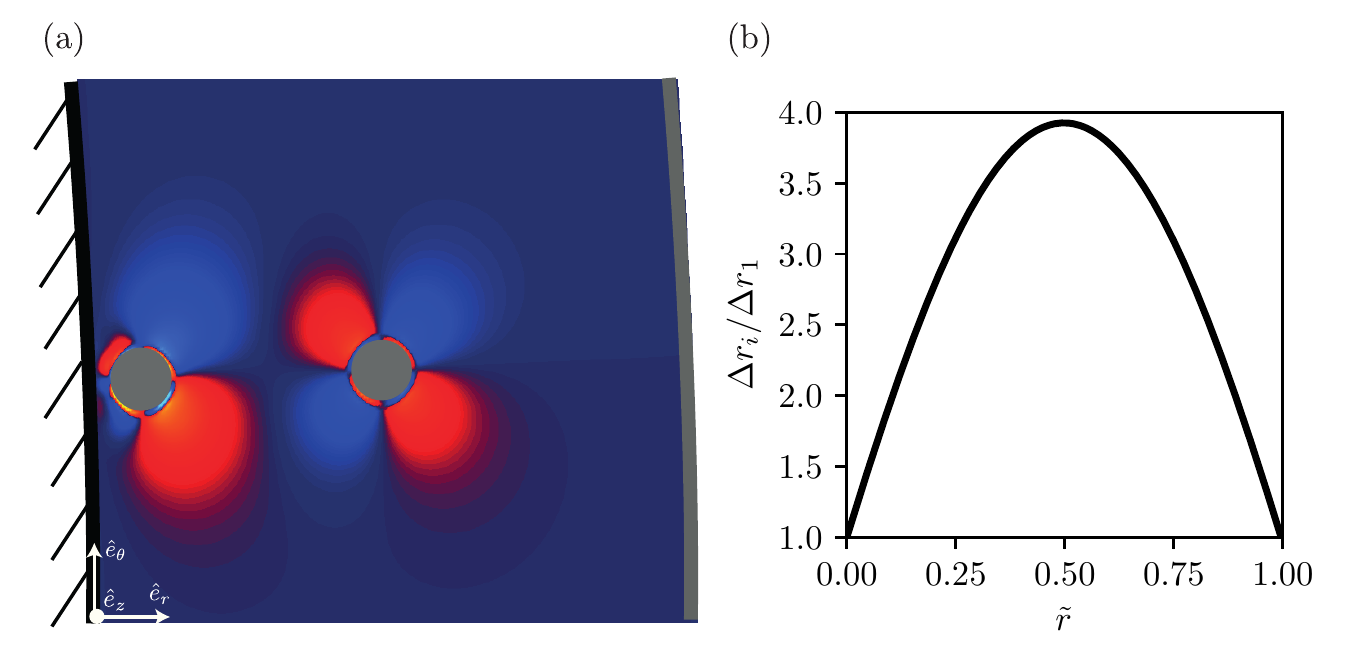}}
  \caption{(a) Snapshot of radial velocity contours in a Taylor-Couette cell with two spherical particles. (b) Grid spacing versus the normalised radial (wall-normal) position}
\label{fig:tcholes}
\end{figure}

We now test this approach for a deforming drop with its centroid fixed immersed in a cross-flow and compute the deformation dynamics arising from the resulting flow conditions with and without refinement. The control parameters are the Reynolds, $Re = Ud_{\text{eq}}/\nu_f$ and the Weber numbers, $We=\rho_fU^2 d_\text{eq}/\sigma$, $d_{\text{eq}}$ is the diameter of the drop in its initial spherical shape. For such a flow, the aspect ratio of the deforming drop depends on the Weber number, which is the ratio of inertia forces acting on the drop in comparison to the surface tension forces. The response of the system can be measured through the quantification of the drop morphology. The combined action of the dynamic pressure and the shear stresses on the surface of the drop leads to its deformation. We adopt a interaction potential based deformation approach geared towards liquid-liquid interfaces as has been described in detail by \citep{spandan2017parallel}. The computational domain is taken of size $L=(10,5,5)d_{\text{eq}}$. The details on the base and refined resolutions of the immersed sphere is given in table \ref{tab:wemesh}. The spherical drop is placed at $\pmb x = (0.5,0.5,0.5)L_z$,  $L_z$ is the height in the vertical direction ($\hat e_z$) and it is bounded by stationary free-slip walls; $\hat e_y$ direction is periodic  and a uniform flow of $\pmb U = U\hat e_x$ is imposed in the $\hat e_x$ direction. In figure \ref{fig:lcomp}(a) we show the morphology of the drop for $We=1.0$ at two different time instants showing its initial spherical state and deformed state. To quantify the shape of the immersed drop we compute the mean aspect ratio of the drop as $X=l_z/l_x$ where $l_z$ and $l_x$ are the lengths of a box surrounding the drop in $\hat e_z$, $\hat e_x$ directions, respectively. In figure \ref{fig:lcomp}(b), we compare the inverse of the mean aspect ratio versus the Weber number for drops using the refinement technique with the simulations without using refinement and also with reference data taken from experimental measurements and a good agreement between all three data-sets is observed. This shows that the adaptive refinement technique can be used successfully to decouple the deformation mesh from the IBM mesh. In order to show the usefulness of this approach we also simulate two stationary bodies, in a Taylor-Couette (TC) cell where the fluid is confined and driven using two independently rotating cylinders. In figure \ref{fig:tcholes}(a) we show two spheres, one close to the wall and the other away from the wall, interacting with a flow being driven in the $\hat e_\theta$ direction. As evidenced in figure \ref{fig:tcholes}(b), the grid is stretched in the radial (wall-normal) direction. One can easily observe that the  flow profiles near the fluid-solid interface are smooth and no discontinuities as seen in figure \ref{fig:fholes}(a) develop in the flow. 

\section{Coarse-grained collision detection}

As mentioned in section 1, when more than one (deformable) body is evolved in a time-dependent chaotic flow, collision among the immersed bodies is unavoidable. Since the enhancements proposed in the paper aims at simulating turbulent flows with a dispersed immiscible phase, an efficient procedure to detect collisions becomes mandatory. The algorithms and models required to satisfactorily detect collisions have received considerable attention in the past given its wide range of applications such as video games, physics based animations, event driven simulations, multiphase flows etc. There exist a number of problem specific techniques in the literature which can be used to detect collision between any two points residing on different objects; see \cite{sigurgeirsson2001algorithms} for a detailed review. The brute force approach is to compute the distance between a marker on one body against every other marker on other bodies to determine a collision event. Using such a simple and straight-forward technique comes at a cost, as the number of operations required to detect a collision becomes terminally expensive with increasing number of markers. Here 'terminally expensive' means an $O(N^2)$ algorithm, where N is the total number of Lagrangian nodes used for all the immersed boundaries. Significant progress has been made in the field of molecular dynamics where the typical approach is to maintain a list of neighbouring particles within a predetermined support radius with whom collision is likely and then update this list periodically throughout the simulation \citep{allen1989computer}. While this a more efficient and computationally inexpensive procedure where the total operation count is  $O(Nlog(N))$ as compared to $O(N^2)$ operations in the brute force method, it still adds additional overhead to the total simulation time per iteration. We propose a computationally inexpensive detection algorithm which makes use of the Eulerian mesh required for the flow solution with minimal overhead. Here, it is important to note that in this section, we focus only on collision detection and not on modelling the collision forces which is an extensive field on its own. 

The basic idea of the algorithm is as follows. A collision event occurs when any two arbitrary Lagrangian markers belonging to two separate immersed bodies occupy the same Eulerian cell. This assumption is reasonable since within the immersed boundary framework the thin film fluid dynamics between two bodies which are just about to collide can never be resolved \citep{thomas2010multiscale}. This problem can be solved by modelling the lubrication forces when the immersed bodies come closer than the width of a single Eulerian cell. Once it is determined that two markers belonging to separate bodies are occupying the same Eulerian cell, a collision force is activated on the respective Lagrangian markers along the normal to the surface of the immersed body. Here it is important to note that the algorithm in particularly useful and computationally efficient when there are several dispersed bodies in the flow which can interact with each other and the surrounding fluid simultaneously.

\begin{figure}
  \centerline{\includegraphics[scale=1.0]{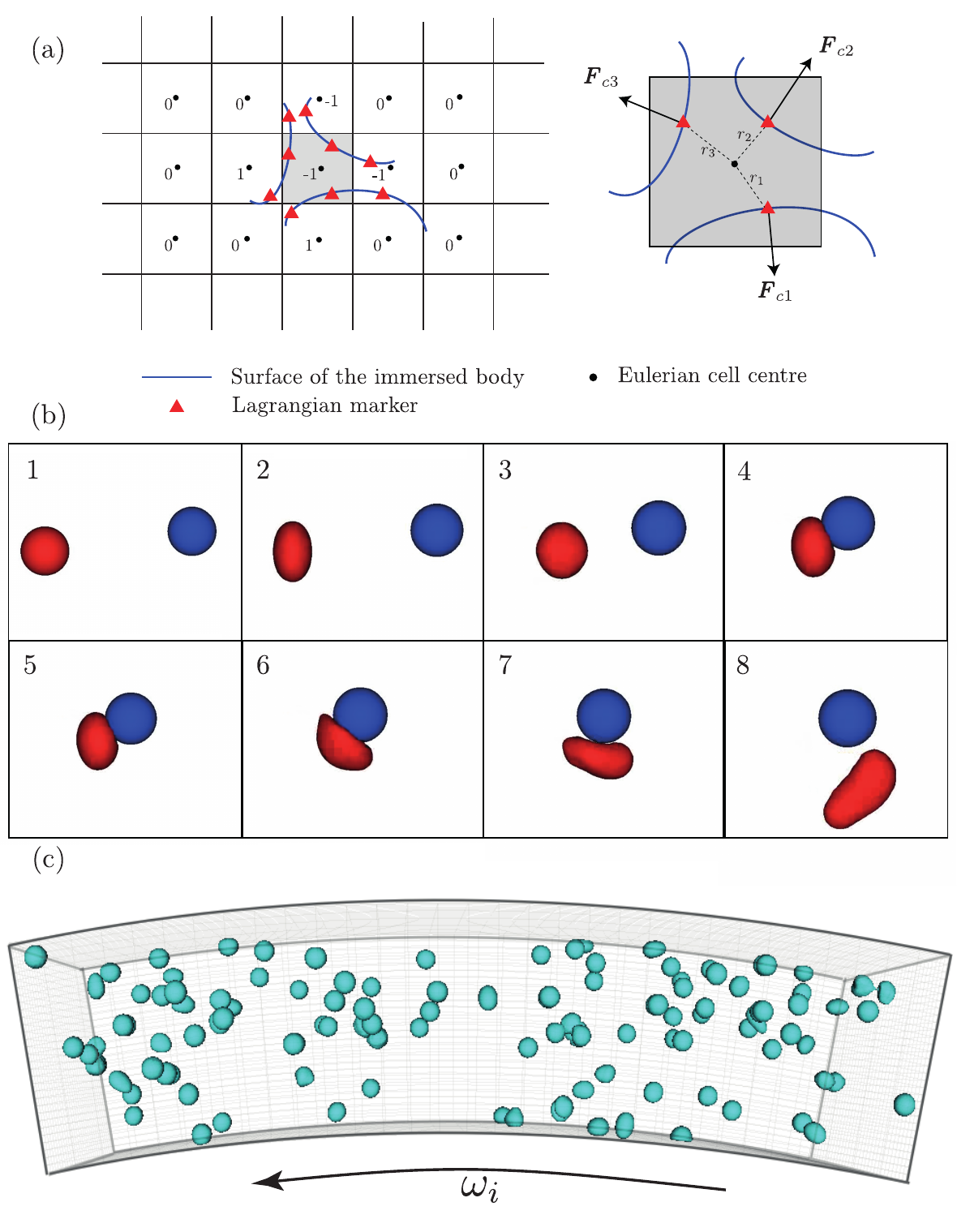}}
  \caption{(a) A schematic of the grid showing the values of the array \texttt{coll} with three different deformable bodies approaching each other in the darkened Eulerian cell. (b) Snapshots at different time instants of a deformable drop (shown in red) driven by a cross flow colliding with a stationary rigid sphere (shown in blue). (c) Instantaneous snapshot of 120 deformable drops interacting with the wall and among themselves in a laminar Taylor-Couette flow.} 
\label{fig:csch}
\end{figure}

First, we initialise a three dimensional integer array \texttt{coll[1:N1,1:N2,1:N3]=0}, where \texttt{N1,N2,N3} are the number of Eulerian grid points in each direction. This array can have three states; \texttt{coll[i,j,k]=0} implies there is no Lagrangian marker occupying the cell, \texttt{coll[i,j,k]>0} implies there are one or more Lagrangian markers belonging to the same immersed body in the cell, \texttt{coll[i,j,k]=-1} implies markers from different immersed bodies occupy the cell and a collision event must occur in this cell. In figure \ref{fig:csch}(a) we show a schematic of an Eulerian grid with three immersed bodies occupying the same Eulerian cell i.e. the value of the array \texttt{coll} in this cell is set to -1. The values of the array \texttt{coll} is shown next to the respective Eulerian cell centres in figure \ref{fig:csch}(a). In such a situation a collision force is activated on each of these markers. For the simulations shown here, we assume that the magnitude of the collision force is inversely proportional to the distance between the Lagrangian marker and the Eulerian centre. The formulation of the force can be of various types where lubrication forces can be included based on the approach velocity of the marker with respect to the Eulerian cell centre; however, the focus here is on the more general problem of detecting a collision event. We now present a pseudo-code for the algorithm which shows its ease of implementation into any general Navier-Stokes solver. 

\begin{verbatim}
coll[1:N1,1:N2,1:N3] = 0

! Routine for enforcing interfacial boundary condition
do pid = 1,N_particle
  do mid = 1,N_f
 	ii = index_of_marker(1,mid,pid)
  	jj = index_of_marker(2,mid,pid)
  	kk = index_of_marker(3,mid,pid)

  	if (coll[ii,jj,kk].eq.0)   coll[ii,jj,kk] = pid
  	if (coll[ii,jj,kk].ne.pid) coll[ii,jj,kk] = -1
  end do
end do

! Routine for computing forces from fluid and internal forces
do pid = 1,N_particle
  do mid = 1,N_f

  	ii = index_of_marker(1,mid,pid)
  	jj = index_of_marker(2,mid,pid)
  	kk = index_of_marker(3,mid,pid)

  	if (coll[ii,jj,kk].eq.-1) -> Activate and compute collision force

  end do
end do
\end{verbatim}

In figure \ref{fig:csch}(b), we show snapshots at different time instants of simulation where a freely moving drop driven by a cross flow collides with a stationary rigid sphere. As can be seen, the collision event occurs smoothly without both bodies penetrating each other. In the supplementary material we have attached an animation which shows collision events between 120 deforming drops with 2562 markers each immersed in a laminar Taylor-Couette flow. An instantaneous snapshot of this animation is shown in figure \ref{fig:csch}(c). Here it is important to emphasize that the overhead in the overall simulation time on implementing the collision-detection algorithm described above is less than 2\% in comparison to other more fine-grained techniques which have a minimum overhead of at least 10\% for such large scale simulations. With the improvements described above, we have been able to simulate turbulent Taylor-Couette flow up to Reynolds numbers of approximately $2\times 10^4$ with hundreds of finite-size bubbles to understand the governing physical mechanisms behind drag reduction \citep{spandan2018physical}.

The next challenge in numerical simulations of finite-size bubbles and drops in turbulent flows is the modelling and implementation of coalescence and breakup phenomena. Within the immersed boundary formulation, both coalescence and breakup events have to be modelled as the interface dynamics are not solved using first principles. This brings into play an array of new challenges as given below, on which we are currently working on. 
\begin{itemize}
\item A strategy to merge / breakup two or more unstructured meshes while ensuring the triangular elements do not distort beyond a critical limit. Here, a choice can be made between (i) stitching the meshes within the numerical simulation which can be numerically expensive for thousands of interacting bubbles/drops (for e.g. \citep{da2014multimaterial,razizadeh2018drop}), or (ii) resort to coarse grained approaches such as importing pre-defined geometries into the simulation during coalescence or breakup .
\item A deterministic strategy to determine when coalescence or breakup occurs. This would depend on several parameters such as approach velocity of the bubbles/drops, their instantaneous shapes, surface tensions etc.  
\item Conservation of mass, momentum and energy during the numerical process of coalescence or breakup. 
\end{itemize}

\section{Summary}
The immersed boundary method (IBM) is becoming a popular tool among computational scientists to tackle problems involving moving and deformable bodies immersed in a fluid. While IBM provides a simple and efficient computational framework for such problems, there are certain inherent issues which limits the scale of the systems affordable. For example, current state of art simulations can handle $O(10^2)$ dispersed deformable bodies in weakly turbulent flows. In this work, we propose and test different algorithms which can be used to handle $O(10^4)$ dispersed deformable bodies in highly turbulent flows within the immersed boundary framework. We describe a fast moving least squares approximation technique where the transfer functions used to exchange information between the Eulerian and Lagrangian meshes are computed only once at the beginning of the simulation and a computationally inexpensive interpolation technique is used during the simulation to estimate the transfer function at any arbitrary point in the domain. We find that with sufficient transfer functions stored at the start of the simulation, we are not only able to estimate the transfer functions within acceptable levels of error but also reduce drastically the total computational time required with increasing immersed Lagrangian markers. We also show that through a systematic refinement of the Lagrangian mesh, it can be decoupled into two meshes; a base mesh which can be used for the solution of the deformation governing equations and a refined mesh which is used to enforce the interfacial boundary condition. Additionally, such a procedure can also be used when the dispersed bodies are immersed in flows with stretched Eulerian grids thus eliminating redundant Lagrangian nodes. Finally, we present a simple and easy to implement algorithm to detect collision events between Lagrangian markers residing on different deformable bodies with a computational overhead of less than 1\% for any number of Lagrangian markers immersed in the flow. 

The authors thank Valentina Meschini for valuable discussions during the course of this work. This work was supported by the Netherlands Center for Multiscale Catalytic Energy Conversion (MCEC), an NWO Gravitation programme funded by the Ministry of Education, Culture and Science of the government of the Netherlands. The simulations were carried out on the national e-infrastructure of SURFsara, a subsidiary of SURF cooperation, the collaborative ICT organization for Dutch education and research. We also acknowledge PRACE for awarding us access to Marconi super computer, based in Italy at CINECA under PRACE project number 2016143351.

\section*{References}

\bibliography{../../mylit}

\begin{thebibliography}{10}
\expandafter\ifx\csname url\endcsname\relax
  \def\url#1{\texttt{#1}}\fi
\expandafter\ifx\csname urlprefix\endcsname\relax\def\urlprefix{URL }\fi
\expandafter\ifx\csname href\endcsname\relax
  \def\href#1#2{#2} \def\path#1{#1}\fi

\bibitem{dowell2001modeling}
E.~Dowell, K.~Hall, Modeling of fluid-structure interaction, Annu. Rev. Fluid
  Mech. 33~(1) (2001) 445--490.

\bibitem{mittal2005immersed}
R.~Mittal, G.~Iaccarino, Immersed boundary methods, Annu. Rev. Fluid Mech. 37
  (2005) 239--261.

\bibitem{peskin1972flow}
C.~S. Peskin, Flow patterns around heart valves: a numerical method, J. Comp.
  Phys. 10~(2) (1972) 252--271.

\bibitem{uhlmann2014sedimentation}
M.~Uhlmann, T.~Doychev, Sedimentation of a dilute suspension of rigid spheres
  at intermediate galileo numbers: the effect of clustering upon the particle
  motion, J. Fluid Mech. 752 (2014) 310--348.

\bibitem{picano2015turbulent}
F.~Picano, W.-P. Breugem, L.~Brandt, Turbulent channel flow of dense
  suspensions of neutrally buoyant spheres, J. Fluid Mechanics 764 (2015)
  463--487.

\bibitem{prosperetti2015life}
A.~Prosperetti, Life and death by boundary conditions, J. Fluid Mech. 768
  (2015) 1--4.

\bibitem{fornari2016sedimentation}
W.~Fornari, F.~Picano, L.~Brandt, Sedimentation of finite-size spheres in
  quiescent and turbulent environments, J. Fluid Mech. 788 (2016) 640--669.

\bibitem{detullio2016moving}
M.~de~Tullio, G.~Pascazio, A {M}oving-{L}east-{S}quares immersed boundary
  method for simulating the fluid-structure interaction of elastic bodies with
  arbitrary thickness, J. Comp. Phys. 325 (2016) 201--225.

\bibitem{spandan2017parallel}
V.~Spandan, V.~Meschini, R.~Ostilla-M{\'o}nico, D.~Lohse, G.~Querzoli, M.~D.
  de~Tullio, R.~Verzicco, A parallel interaction potential approach coupled
  with the immersed boundary method for fully resolved simulations of
  deformable interfaces and membranes, J. Comp. Phys. 348 (2017) 567--590.

\bibitem{iaccarino2003immersed}
G.~Iaccarino, R.~Verzicco, Immersed boundary technique for turbulent flow
  simulations, App. Mech. Rev. 56~(3) (2003) 331--347.

\bibitem{schwarz2016immersed}
S.~Schwarz, T.~Kempe, J.~Fr{\"o}hlich, An immersed boundary method for the
  simulation of bubbles with varying shape, J. of Comp. Phys. 315 (2016)
  124--149.

\bibitem{uhlmann2005immersed}
M.~Uhlmann, An immersed boundary method with direct forcing for the simulation
  of particulate flows, J. Comp. Phys. 209~(2) (2005) 448--476.

\bibitem{vanella2009moving}
M.~Vanella, E.~Balaras, A moving-least-squares reconstruction for
  embedded-boundary formulations, J. Comp. Phys. 228~(18) (2009) 6617--6628.

\bibitem{belytschko1994fracture}
T.~Belytschko, L.~Gu, Y.~Lu, Fracture and crack growth by element free galerkin
  methods, Modelling and Sim. in Mat. Sci. and Engg. 2~(3A) (1994) 519.

\bibitem{belytschko1996meshless}
T.~Belytschko, Y.~Krongauz, D.~Organ, M.~Fleming, P.~Krysl, Meshless methods:
  an overview and recent developments, Computer Methods in App. Mech. and Engg.
  139~(1) (1996) 3--47.

\bibitem{krongauz1996enforcement}
Y.~Krongauz, T.~Belytschko, Enforcement of essential boundary conditions in
  meshless approximations using finite elements, Computer Methods in App. Mech.
  and Engg. 131~(1) (1996) 133--145.

\bibitem{hegen1996element}
D.~Hegen, Element-free {G}alerkin methods in combination with finite element
  approaches, Computer Methods in App. Mech. and Engg. 135~(1) (1996) 143--166.

\bibitem{atluri1999analysis}
S.~Atluri, J.~Cho, H.-G. Kim, Analysis of thin beams, using the meshless local
  {P}etrov--{G}alerkin method, with generalized moving least squares
  interpolations, Computational Mechanics 24~(5) (1999) 334--347.

\bibitem{schaefer2006image}
S.~Schaefer, T.~McPhail, J.~Warren, Image deformation using moving least
  squares, in: ACM Trans. on Graphics, Vol.~25, ACM, 2006, pp. 533--540.

\bibitem{fleishman2005robust}
S.~Fleishman, D.~Cohen-Or, C.~Silva, Robust moving least-squares fitting with
  sharp features, in: ACM Trans. on Graphics, Vol.~24, ACM, 2005, pp. 544--552.

\bibitem{kolluri2008provably}
R.~Kolluri, Provably good moving least squares, ACM Trans. on Algorithms 4~(2)
  (2008) 18.

\bibitem{zeng2004curve}
Q.-H. Zeng, D.-T. LU, Curve and surface fitting based on moving least-squares
  methods, J. of Engg. Graph. 1~(3) (2004) 84--87.

\bibitem{kobbelt2004survey}
L.~Kobbelt, M.~Botsch, A survey of point-based techniques in computer graphics,
  Comp. \& Graph. 28~(6) (2004) 801--814.

\bibitem{liu2005introduction}
G.~Liu, Y.~Gu, An introduction to meshless methods and their programming,
  Springer, Berlin, Germany, 2005.

\bibitem{liu2003smoothed}
G.-R. Liu, M.~B. Liu, Smoothed particle hydrodynamics: a meshfree particle
  method, World Scientific, 2003.

\bibitem{kempe2015imposing}
T.~Kempe, M.~Lennartz, S.~Schwarz, J.~Fr{\"o}hlich, Imposing the free-slip
  condition with a continuous forcing immersed boundary method, J. Comp. Phys.
  282 (2015) 183--209.

\bibitem{abraham1970functional}
F.~F. Abraham, Functional dependence of drag coefficient of a sphere on
  {R}eynolds number, Phys. Fluids 13~(8) (1970) 2194--2195.

\bibitem{loth2008quasi}
E.~Loth, Quasi-steady shape and drag of deformable bubbles and drops, Int. J.
  Multi. Flow 34~(6) (2008) 523--546.

\bibitem{sigurgeirsson2001algorithms}
H.~Sigurgeirsson, A.~Stuart, W.-L. Wan, Algorithms for particle-field
  simulations with collisions, J. Comp. Phys. 172~(2) (2001) 766--807.

\bibitem{allen1989computer}
M.~Allen, D.~Tildesley, Computer simulation of liquids, Oxford university
  press, 1989.

\bibitem{thomas2010multiscale}
S.~Thomas, A.~Esmaeeli, G.~Tryggvason, Multiscale computations of thin films in
  multiphase flows, Int. J. Multiphase Flow 36~(1) (2010) 71--77.

\bibitem{spandan2018physical}
V.~Spandan, R.~Verzicco, D.~Lohse, Physical mechanisms governing drag reduction
  in turbulent {T}aylor-{C}ouette flow with finite-size bubbles, J. Fluid Mech.
  (under review).

\bibitem{da2014multimaterial}
F.~Da, C.~Batty, E.~Grinspun, Multimaterial mesh-based surface tracking., ACM
  Trans. Graph. 33~(4) (2014) 112--1.

\bibitem{razizadeh2018drop}
M.~Razizadeh, S.~Mortazavi, H.~Shahin, Drop breakup and drop pair coalescence
  using front-tracking method in three dimensions, Acta Mechanica 229~(3)
  (2018) 1021--1043.

\end{thebibliography}

\end{document}